\documentstyle[aps,prl,multicol]{revtex}
\begin{document}

\title{Universal Relationship Between Giant Magnetoresistance and Anisotropic
Magnetoresistance in Spin Valve Multilayers}
\author{B.\ H.\ Miller\cite{brad_seagate}$^1$, Branko P.\ Stojkovi\'c,$^2$
and E.\ D.\ Dahlberg$^1$}
\address{$^1$University of Minnesota, School of Physics and Astronomy,
Minneapolis, MN 55455\\
$^2$Center for Nonlinear Studies, Los Alamos National Laboratory,
Los Alamos, NM 87545}
\maketitle

\begin{abstract}
\leftskip 54.8pt
\rightskip 54.8pt
We measure the giant magnetoresistance (GMR) with the current
both parallel and perpendicular to the
direction of the magnetization in the ferromagnetic (FM)
layers and thus probe the anisotropy of the
effective mean free paths for the spin-up
and spin-down electrons, seen in the
anisotropic magnetoresistance. We find that the
difference of the GMR in the two configurations, when expressed in
terms of the sheet conductance, displays a nearly universal behavior
as a function of GMR.  On interpreting the results within 
the Boltzmann transport formalism we demonstrate 
the importance of bulk scattering for GMR.
\end{abstract}

\begin{multicols}{2}
The giant magnetoresistance (GMR) occurs in multilayers
consisting of adjacent layers of FM and
non-FM materials\cite{Binasch,Dupas} and is characterized by the
large change in conductivity as the magnetic moments
in the adjacent FM layers change from parallel to antiparallel.
Experimentally, it has been determined that the amplitude
of GMR depends strongly on the thickness of the
constituent layers\cite{Dieny}
and on the properties of the interfaces between
them\cite{Parkin,Ivan}.
While there is a consensus
that the spin-dependent scattering, present in FM
materials \cite{MottA,MottB,Fert,Gurney}, is responsible for the
effect\cite{Camley,Levy,Barth,Edwards,Barnas},
it is still unclear whether the interfacial scattering
or bulk scattering plays a more important role.

In this letter we report the results of a systematic study, aimed at
sorting out the relative importance of the interfacial and bulk
scattering processes.  To accomplish this, we measure differences in
two GMR configurations, one with the applied current {\em parallel}
($GMR_{{\vert\vert}}$) and the second with the current {\em
  perpendicular} ($GMR_{\perp}$) to the magnetization of the sample.
We find $GMR_{\perp} > GMR_{{\vert\vert}}$ for all 52 samples in this
study.  We attribute the difference in the two GMR values to the change
in the difference in the mean free paths of spin-up and spin-down
electrons in the two current/magnetization orientations.  Since the
current/magnetization orientation effect on the mean free paths of the
spin-up and spin-down electrons, a phenomenon well known as the
anisotropic magnetoresistance (AMR), is a bulk effect \cite{McGuire},
we find that the GMR is also dominated by bulk spin-dependent
scattering.

The films used in this study have been prepared by d.c.\  magnetron
sputtering on Si(100)/SiN$_{x}$(100nm) substrates, patterned for lift-off
photo-lithography. They consist of two FM
layers (made of Co and/or NiFe), separated by a non-magnetic (Cu or CuSn)
spacer.
A total of 52 samples, which can be divided into six
series, have been measured (see Table 1).
The top Co layer is allowed to partially oxidize in the air, creating a
thin antiferromagnetic (AF) layer of CoO,
with a bulk N\'eel temperature $T_N$ of $\sim 290$ K.
Magnetic moment measurements, obtained using a Quantum Design SQUID
magnetometer, indicate that the top Co layer thickness
is typically reduced by $t_{ox}=1-1.5$ nm by oxidation\cite{details}.
When the samples are cooled
to $T=4.2K\ll T_N$ in a magnetic field of 2000 Oe, the AF
layer provides a unidirectional exchange anisotropy, which pins
the direction of the magnetization in the
adjacent (top) FM layer, while in the bottom FM layer the
magnetization is free to follow an applied field\cite{spin-valves}.
The orientation of the exchange anisotropy is easily altered by
thermal cycling and resetting the applied field
direction before cooling again.

We measure the GMR in both parallel and perpendicular configurations 
using the standard
four-point probe technique with a {\em well defined} 
(1mm x 10mm) conduction path\cite{dieny_4}.
We express the measured results in terms of sheet conductance,
$G=\sigma t$, where $\sigma$ is the conductivity and $t$ is the
film thickness. The
four quantities necessary to describe the
experiment are denoted by $G^{\uparrow\uparrow}_{{\vert\vert}}$,
$G^{\uparrow\downarrow}_{{\vert\vert}}$, $G^{\uparrow\uparrow}_{\perp}$, and $G^{\uparrow\downarrow}_{\perp}$,
where $||$ and $\perp$ refer to the orientation of the magnetization
relative to the current direction and $\uparrow$ and $\downarrow$ refer to the
orientation of the magnetization in the ferromagnetic layers.
Hence we define
\begin{mathletters}
\begin{equation}
GMR_{{\vert\vert}} = G^{\uparrow\uparrow}_{{\vert\vert}} - G^{\uparrow\downarrow}_{{\vert\vert}} ,
\qquad GMR_{\perp} = G^{\uparrow\uparrow}_{\perp} - G^{\uparrow\downarrow}_{\perp} ,
\end{equation}
and
\begin{equation}
\Delta GMR = GMR_{\perp} - GMR_{{\vert\vert}} .
\end{equation}
\end{mathletters}
Figure \ref{fig1} shows
the sheet conductance in a typical sample
at $T=4.2$K for all states of interest:
clearly, the four quantities are well defined, i.e., the energy
of the exchange anisotropy of the pinned FM layer is high enough to provide
a signal which is stable with increasing field strength.
More importantly, the results show that the $GMR$ is larger in
the perpendicular (higher conductance AMR) state.
If one assumes that a Drude-like
formalism can be applied, then this fact alone implies that the GMR is
a probe of the difference of the mean free paths $\lambda$
of the spin-up ($\lambda^{\uparrow}$) and spin-down ($\lambda^{\downarrow}$)
electrons ($\lambda^{\uparrow}\gg\lambda^{\downarrow}$).
In general, films with larger values of $GMR$ are more
sensitive to the change in the differences in mean free paths,
caused by changing the
magnetization from parallel to perpendicular relative
to the current direction.

In Figure 2, which is
the central result of this paper, we plot $\Delta GMR$
as a function of $GMR_{{\vert\vert}}$ (see
Eq.\  (1)) in comparison with our  calculated results (see
below).  The experimental
data, plotted in terms of the {\em sheet conductance},
show striking universality up to rather large values of $GMR_{{\vert\vert}}$.
Moreover the universality is not constrained only to different
geometries of Co/Cu/Co, the primary trilayer studied here,
but is also applicable to different materials (see Table 1, series III and
VI), as well as samples without a Ta seed layer (series IV).
We emphasize that the {\em relative} change in $GMR$, defined as
$GMR_{\perp,||}(\%) =1-G^{\uparrow\downarrow}_{\perp,||}/G^{\uparrow\uparrow}_{\perp,||}$
does not provide a universally behaved $\Delta GMR(\%)\equiv
GMR_{||}(\%)-GMR_{\perp}(\%)$, as seen in the inset of Figure 2.
Hence, our result supports the view that
the change in the sheet conductance, $G$, is the fundamental
measure of GMR\cite{Nozieres}.

In order to eliminate possible
systematic errors in the measurement technique
we have performed a number of checks. For example,
by replacing the
Cu layer with a binary alloy
Cu$_{98}$Sn$_{2}$ (series III in Table 1)
whose bulk resistivity at 4.2K is
$\sim$6.2 $\mu\Omega$ cm which is
somewhat {\em larger} than that in Co
($\sim 5\, \mu\Omega$cm)
and much larger than that of Cu, we have verified
that the behavior of $\Delta GMR$ as a function of $GMR$ is not due to
current shunting by the copper, rather than the aforementioned change in
the mean free paths. As seen in Figure 2, in this case
the form of $\Delta GMR$ is virtually identical
to that of films with a thin Cu spacer.
Moreover, with the presumption that the interfaces in these
polycrystalline samples are not atomically smooth,
we can exclude internal reflection in the Cu\cite{butler}
as a possible explanation of the observed $\Delta GMR$.
We have also examined whether the thermal cycling causes a shift in the
base resistance, which could produce a non-vanishing value of $\Delta GMR$
observed in the experiment: we compared the value of $AMR = \rho_{||}
- \rho_{\perp}$, where $\rho_\perp$ and $\rho_{||}$
are obtained in two thermal cycles, to the AMR
measured by rotating the magnetization in
both FM layers in a large magnetic field
(10kOe).  In the latter case the resistivity was measured as a function of
the angle $\theta$ between ${\bf j}$ and ${\bf M}$
{\bf (}i.e., $AMR=\rho(\theta=0)-\rho(\theta=\pi/2)${\bf )}.
This comparison yielded virtually no difference between
the two values of the AMR and hence no systematic shifts in GMR due to
thermal cycling.

For the analysis of our results, we use the
semi-classical approach to GMR\cite{Camley,Barnas,Dieny-jphys}.
This approach is similar to that of Rijks et al\cite{rijks}, although
here we allow for spin dependent scattering at
interfaces.
Starting from the
Boltzmann equation and neglecting deviations in Ohm's law, the displacement
of the fermionic distribution function, $g({\bf v},z)=f - f_{o}$,
is given by
\begin{equation}
g^s_{i\pm} =
{ e\tau^{s}_{i} E \over m^{s}_{i}}
\left({\partial f_{0} \over \partial v_{x}}\right)
\left[ 1-A^{s}_{i\pm}({\bf v})
\exp\left( {\mp z \over \tau^{s}_{i}\vert v_{z}\vert}
\right)
\right]
\label{eq3}
\end{equation}
where $e$ and $m$ are the electron charge and mass respectively, $v_{z}$
is the
electron velocity perpendicular to the film plane, $\tau$ is the mean time
between scattering events, $f$ is
the steady state Fermi distribution, and $f_{o}$
is the equilibrium Fermi distribution.  The functions $A^{s}_{i\pm}({\bf v})$
are obtained from the following boundary conditions:
\begin{equation}
g^{s}_{1+}(z=0,v_{z}) = p\, g^{s}_{1-} (z=0,v_{z})
\end{equation}
\begin{equation}
g^{s}_{3-}(z=d,v_{z}) = q\, g^{s}_{3+} (z=d,v_{z})
\end{equation}
and
\begin{equation}
g^{s}_{(i\pm 1)\pm}(z_{i},v_{z}) = T^{s}_{(i\rightarrow i\pm 1)}\,
g^{s}_{i\pm} (z_{i},v_{z})
\end{equation}
where ${\bf v}$ is the quasiparticle velocity,
$p$ and $q$ represent the fraction of electrons which are specularly
scattered from the bottom and top surfaces respectively
and $T^s_{(i\rightarrow i\pm 1)}$ are the transmission coefficients.
The subscripts $\pm$ indicate whether an electron is
traveling in the positive or negative
$z$ direction, and the superscript $s$ represents the
spin state with respect to the magnetization in a FM layer.

From Eqs.\ (2)-(5) the sheet conductivity is straightforwardly obtained using
\begin{equation}
G = {J_{x} t \over E} =
{-2e \over E}\left({m \over h}\right)^3\int_{0}^{t}dz
\int_{-\infty}^{\infty} d^{3} v\,v_{x}\, g({\bf v},z)
\label{eq:j}
\end{equation}
where $v_{x}$, $E_x$ and $J_{x}$ are the $x$-components of the
velocity, applied electrical field and current respectively.
It is easily verified that the ``bulk'' sheet conductance
of the individual layers, $\sigma t_i$, where $t_i$ is the
thickness of $i$th layer, is subtracted out
and does not contribute to the $GMR$\cite{details}.

Equation (6) is most efficiently solved numerically.
Clearly there are many parameters in
the model, however, several experimental constraints can limit the
parameter space considerably.  For example, studies of the effect of
surface scattering in polycrystalline thin films show that $p$ and $q$
are nearly zero\cite{Mayadas}.
In addition, our measured results show no substrate/seed layer
dependence of $\Delta GMR$ vs $GMR$ and an explicit
calculation shows that for $p$, $q<0.5$
our numerically obtained $\Delta GMR$ vs $GMR$
depends very little on $p$ and $q$. Hence we set $p=q=0$.
In this limit one can obtain $GMR$ analytically
and express the result in terms of exponential integrals, $E_1(x)$,
where the argument $x$ depends only on {\em ratios} of thicknesses and
mean free paths\cite{details}.

The coefficients $T^s_{(i\rightarrow i\pm 1)}$
represent the fraction of electrons specularly
transmitted through an interface, with
the subscript $i$ corresponding to the layer from which the
electron was emitted.  The remaining electrons are
either diffusively transmitted or diffusively reflected, and in
principle one should include the specular reflection coefficient $R$ into the
calculation.  However, in a polycrystalline sample, with presumably
rough interfaces, electrons reflected at an interface
will predominantly scatter diffusively and hence we set $R = 0$.
Although, strictly speaking, $T^s_{(i\rightarrow i\pm 1)}$
depend on both $i$ and $s$,
here we assume that $T^s_{(i\rightarrow i\pm 1)}$ are the same at
two interfaces; that is, we assume that the transmission
coefficients do not depend on the incident layer
($T^s_{(1\rightarrow 2)}=T^s_{(2\rightarrow 1)}=T^s_{(2\rightarrow 3)}=
T^s_{(3\rightarrow 2)}\equiv T^s$).  Thus the eight transmission coefficients
have been reduced to $T^\uparrow$ and $T^\downarrow$.

Within this model the
observed $\Delta GMR$ is caused by the angular anisotropy of
either the mean free paths or the transmission coefficients.
Hence we extend the work of Ref.\ \onlinecite{Dieny-jphys}
and assume that in the magnetic layers (corresponding to indices $i=1,3$)
$\lambda^s_i$ depend on the angle $\theta$ between ${\bf M}$ and
${\bf j}$ as in AMR, while in the non-magnetic interlayer the mean free
path has neither a spin nor angular dependence,
$\lambda_2(\theta)=\lambda_2=const$.
In principle, the transmission
coefficients $T^s$ may exhibit anisotropy as well\cite{stiles}; however,
in the polycrystalline samples studied here this anisotropy is
averaged to zero.
Moreover, it is easy to verify, using the
aforementioned analytical solution of Eq.\ (\ref{eq:j}),
that the angular
change in $T^s$, $\Delta T^s = T^s_{||} -T^s_{\perp}$,
yields $\Delta GMR\propto (\Delta T^s/T^s)  GMR$ and no higher order
terms, regardless of whether $GMR$ is due to the spin-dependent
mean free paths or transmission coefficients, i.e.,
$\Delta GMR$ plotted against $GMR$ would yield a straight line. 
Hence,  the experimentally observed
curvature of $\Delta GMR$ vs $GMR$,  seen in Fig.\ 2,
is in clear contradiction with $\Delta GMR$ being due to
the anisotropy of $T^s$, i.e., within this formalism the observed
$\Delta GMR$ is due to the angular anisotropy of the conduction electron mean
free paths in the ferromagnetic layers (the origin of the AMR).
Indeed, with reasonable values of $\lambda^\uparrow_{\vert\vert}$ and
$\lambda^\uparrow_\perp$, estimated from the the bulk resistivity anisotropy
of Co at $T=4.2$ K, one can easily fit the data 
(see the solid line in Fig.\ 2).

Analytically, it is straightforward to show that, both for
$\lambda^\uparrow\gg\lambda^\downarrow$ and/or $T^\uparrow >
T^\downarrow$, one has $\Delta GMR\propto (\Delta
\lambda^s/\lambda^s)\, GMR$ plus higher order (in $GMR$) terms. Hence,
the universal slope of $\Delta GMR$ at low $GMR$ indicates that the
bulk magnetic scatterers (responsible for the AMR) dominate scattering
in both magnetic materials used here (Co and NiFe). In addition, for
thin non-FM spacer our theoretical solution for $\Delta GMR$ does not
depend on $\lambda_2$, in agreement with experiment. However, our
experimental results show that $\Delta GMR$ as a function of $GMR$ is
independent of both $\lambda^s$ and $T^s$. Both analytically and
numerically one can show that this is possible only if
$T^\uparrow\approx T^\downarrow$\cite{details}. Moreover, the
sensitivity of the calculated result to the change in
$T^\uparrow-T^\downarrow$ can be relatively large, and if the GMR is
due to the spin-dependent transmission coefficients, then the
experiment would likely show a non-monotonic behavior of $\Delta GMR$
vs $GMR$ even within the same series.  Thus the observed universal
behavior of $\Delta GMR$ implies that $T^s$ are only weakly spin
dependent.  From the amount of scatter in Fig.\ 2 we find that the
difference between $T^\uparrow$ and $T^\downarrow$ is at most about
10\%.

The measured  result for series II  requires further explanation:
in samples with $t_{Cu}>4$ nm $\Delta GMR$  as a function of $GMR$
deviates from the otherwise universal behavior (see Fig.\ 2).
Moreover, in the same series we find that $GMR$, 
expressed in terms of the sheet conductance (1), 
{\em increases} with increasing Cu thickness up to $t_{Cu}=10$ nm,
which is physically counter intuitive.
However, the present model assumes that the mean free 
path in Cu does not depend on the Cu layer thickness. It is
well known that $\lambda_{Cu}$  depends on
the grain size and that the grain sizes increase
with increasing film thickness. Therefore it is reasonable to assume
that $\lambda_{Cu}$ increases with increasing Cu thickness.
We have verified by explicit calculation that this indeed 
yields the experimentally observed result \cite{details}.

In conclusion, we have performed a systematic examination
of the GMR effect in
spin-valves in two configurations, with the applied current
parallel and perpendicular to the magnetization direction.  The obtained
results are in agreement with the fact that the sheet conductance
is the most fundamental measure of the magneto-transport in thin
films and multilayers. We find that the difference of $GMR$ in
the two configurations, $\Delta GMR$, is a universal function of $GMR$
and is due to the
angular dependence of the electronic mean free path for spin-up
and spin-down electrons.  We argue on physical grounds that
the spin dependent transmission coefficients cannot yield
the observed behavior. The universality of the behavior of
$\Delta GMR$ vs $GMR$, both as a function of layer thickness and
the mean free paths, demonstrates that bulk magnetic scatterers
are primarily responsible for the GMR.

We wish to thank A. Balatsky for useful conversations.
This work has been supported by the U.S.\ Department of Energy
and by the AFOSR under grant no. AF/FA9620-92-J-0185.

\end{multicols}

\begin{table}
\caption{Thickness (in nm) of the various layers in the samples.  There are
a total of 52 films, separated into six series.  In Series V an
attempt was made to keep the two Co layers of equal thickness after
oxidation of the upper Co layer.  Note that films with NiFe
in Series VI contain a 0.5nm layer of Co deposited on the exposed
surface, necessary to provide the AF layer of CoO.}
\begin{tabular}{|c|c|c|c|c|c|c|c|c|}
Series & Ta & Ni$_{81}$Fe$_{19}$ & Co &
Cu & Cu$_{98}$Sn$_2$ & Co & Ni$_{81}$Fe$_{19}$ & Co \\
\hline
I & 2.5 & 0 & 4-20 & 2.5 & 0 & 4 & 0 & 0 \\
\hline
II & 2.5 & 0 & 5 & 2-10  & 0 & 4 & 0 & 0 \\
\hline
III & 2.5 & 0 & 2.5-20 & 0 & 2.3 & 4 & 0 & 0 \\
\hline
IV & 0 & 0 & 2-20 & 2.5 & 0 & 4 & 0 & 0 \\
\hline
V & 0, 5 & 0 & 2.7-10 & 2.1-2.5 & 0 & 4.5-11 & 0 & 0 \\
\hline
VI & 0 & 2.5, 5 & 0, 0.5 & 2.5 & 0 & 0, 0.5 & 0, 2.5, 5 & 0.5 
\end{tabular}
\end{table}

\begin{figure}
\caption{Typical sheet conductance vs applied field for the magnetization
parallel (open circles) and perpendicular (filled circles) to the applied
current.  The data are obtained in Ta 2.5nm/Co 5nm/Cu 2.5nm/Co (4-$t_{ox}$)
nm film.  For $H < 0$ the magnetization of the two layers are parallel
($\uparrow\uparrow$) and for $H > 0$ and they are antiparallel 
($\uparrow\downarrow$).  The GMR
corresponds to $G^{\uparrow\uparrow} - 
G^{\uparrow\downarrow}$ with the applied
current  either parallel (${\vert\vert}$) or 
perpendicular ($\perp$) to the sample
magnetization. The data, together with SQUID magnetization
measurements, 
indicate that the magnetic states in Co layers are well defined.}
\label{fig1}
\end{figure}

\begin{figure}
\caption{$\Delta GMR = GMR_{\perp} - GMR_{\vert\vert}$
  versus $GMR_{\vert\vert} = G_{\vert\vert}^{\uparrow\uparrow}-
  G_{\vert\vert}^{\uparrow\downarrow}$, for the series depicted in
  Table 1, where $G_{{\perp}
    (||)}^{\uparrow\uparrow(\uparrow\downarrow)}$ is the sheet
  conductance measured in magnetically parallel (antiparallel) states
  and with the applied current perpendicular (parallel) to the sample
  magnetization; an estimated relative error is less than 1\%. 
  The data display a general trend for all films,
  independent of material with only a deviation for
  the thickest films in series II.  The solid line shows the
  calculated result, obtained with $\lambda_{||}^{\uparrow}=30$ nm,
  $\lambda^\downarrow=3$ nm, $\lambda_\perp^\uparrow -
  \lambda_{||}^\uparrow = 1.2$ nm and $T^\uparrow=T^\downarrow=0.5$
  which corresponds to the GMR being due to bulk spin-dependent
  scattering.  $\Delta GMR$ calculated using anisotropic transmission
  coefficients is always linear in $GMR$, as depicted by the dashed
  line, obtained with $\lambda_{(||,\perp)}^{s}=30$ nm,
  $T^\uparrow_\perp=0.8$, $T^\uparrow_{||}=0.81$ and $T^\downarrow =
  0.4$. In all calculations we have assumed that the mean free path in
  Cu is $\lambda_{Cu}=50$ nm.  The thickness of the various layers
  were chosen directly from Table 1. Inset: $\Delta GMR$ vs the
  relative change in GMR (with $GMR$ given in percent), for films
  depicted in Table 1, displays a non-universal behavior.  }
\end{figure}


\begin{references}
\bibitem{details} Details of this study will be given elsewhere.
\bibitem{spin-valves} Films of this type are
frequently referred to as ``spin-valves.''
\end{references}
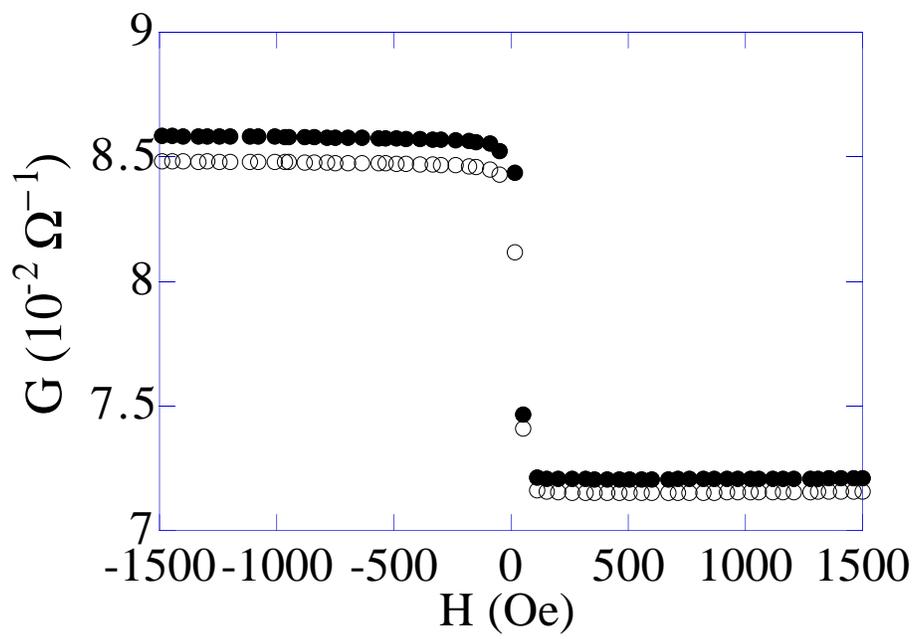
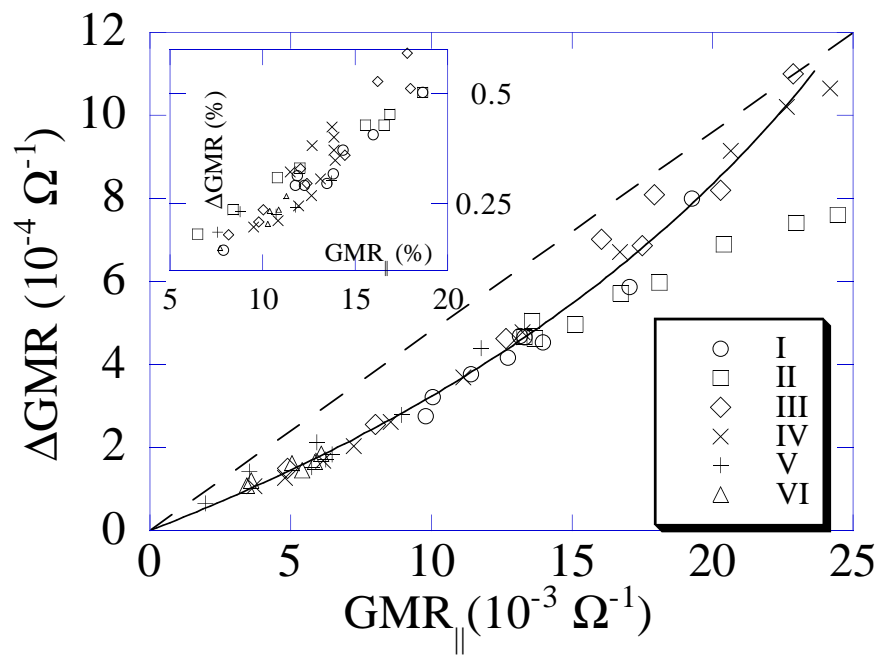
\end{document}